\documentclass{aa}  

\usepackage{graphicx}
\usepackage{txfonts}
\usepackage{hyperref}
\hypersetup{
    colorlinks=true,
    allcolors=blue,
    }
\usepackage{graphicx}
\usepackage{dcolumn}
\usepackage{bm}
\usepackage{xcolor}
\usepackage{tikz}
\usetikzlibrary{intersections}
\usepackage[utf8]{inputenc}
\usepackage[T1]{fontenc}
\usepackage{etoolbox}
\usepackage{mathtools}
\usepackage{float}
\usepackage{physics}
\usepackage{subfig}
\usepackage{tabularx,ragged2e,booktabs,caption}
\usepackage{natbib}
\bibpunct{(}{)}{;}{a}{}{,}
\DeclareMathAlphabet{\mathcal}{OMS}{cmsy}{m}{n}
\begin{document}

   \title{Relativistic shocks in conductive media}


   \author{Argyrios Loules
          \and
         Nektarios Vlahakis
          }

   \institute{Section of Astrophysics, Astronomy and Mechanics, Physics Department, National and Kapodistrian University of Athens, Panepistimiopolis, 15784 Zografos, Athens, Greece\\
              \email{arloules@phys.uoa.gr, vlahakis@phys.uoa.gr}
         }

   \date{}

 
  \abstract
   {Relativistic shocks are present in all high-energy astrophysical processes involving relativistic plasma outflows interacting with their ambient medium. While they are well understood in the context of relativistic hydrodynamics and ideal magnetohydrodynamics (MHD), there is a limited understanding of the properties related to their propagation in media characterized by finite electrical conductivity.}
   {This work presents a systematic method for the derivation and solution of the jump conditions for relativistic shocks propagating in MHD media with finite electrical conductivity. This method is applied to the numerical solution of the Riemann problem and the determination of the conditions inside the blastwave that is formed when ultrarelativistic magnetized ejecta interact with the circumburst medium during a gamma-ray burst.}
   {We derived the covariant relations expressing the jump conditions in a frame-independent manner. The resulting algebraic equations expressing the Rankine-Hugoniot conditions in the propagation medium's frame were then solved numerically. A variable adiabatic index equation of state was used in order to obtain a realistic description of the post-shock fluid's thermodynamics. This method was then employed for the solution of the Riemann problem for the case of a forward and a reverse shock, both of which form during the interaction of a gamma-ray burst ejecta with the circumburst medium. This allowed us to determine the kinematics of the resulting blastwave  and the dynamical conditions in its interior.}
   {Our solutions clearly depict the impact of the plasma's electrical conductivity in the properties of the post-shock medium. Two characteristic regimes are identified with respect to the value of a dimensionless parameter that has a linear dependence on the conductivity. For small values of this parameter, the shock affects only the hydrodynamic properties of the propagation medium and leaves its electromagnetic field unaffected. No current layer forms in the shock front; thus, we refer to this as the current-free regime. For large values of this parameter, the ideal MHD regime has been retrieved. We also show that the assumption of a finite electrical conductivity can lead to higher efficiencies in the conversion of the ejecta energy into thermal energy of the blastwave through the reverse shock. The theory developed in this work can be applied to the construction of Riemann solvers for resistive relativistic MHD (RRMHD).}
   {}


   \maketitle
%

\section{Introduction}

Shocks form naturally in fluid dynamics through the nonlinear evolution of waves and they are especially important  when a perturbation propagates in a medium with velocity greater than the fastest local wave velocity. The medium's density, pressure, bulk velocity, and, in the case of magnetized plasma, its electromagnetic field, exhibit steep changes or jumps across the shock front \citep{pain1963}. The investigation of the dynamics of both Newtonian and especially relativistic shocks propagating in hydrodynamic and magnetohydrodynamic (MHD) media has been a field of particular interest, due to their driving role in energetic astrophysical phenomena, such as gamma-ray bursts (GRBs), internal shocks in active galactic nucleus (AGN) jets, and the interaction of relativistic pulsar winds with their host nebulae \citep{bykov2011}, for instance. In GRBs in particular, the forward and reverse shocks that appear at the onset of the ejecta-circumburst medium (CBM) interaction \citep{sari1995, zhang2005} shape the early phase of the observed afterglow emission \citep{mimica2010, gao2015}.

The properties of relativistic shocks have been studied extensively, notably by \citet{taub1948}, who derived the Rankine-Hugoniot conditions for hydrodynamic relativistic shocks, as well as \citet{dehoffman1950}, who investigated the dynamics of relativistic shocks in magnetized plasmas. \citet{bm1976}, following Taub's method, presented solutions to the jump conditions for a strongly relativistic shock propagating in a cold, hydrodynamic medium, expressed in the undisturbed fluid's comoving frame. They also modelled the dynamics of spherically symmetric relativistic blast waves through self-similar solutions. The jump conditions for strongly relativistic shocks in magnetohydrodynamic media were later derived by \citet{kennel1984}. The theory of relativistic shocks has also been applied to the determination of the conditions inside the blastwave in the early phases of the ejecta-CBM interaction. Under the assumption of a uniform flow Lorentz factor and total pressure in the region between the forward shock (FS) and the reverse shock (RS), we can solve the jump conditions for both shocks and determine the pressure and velocity of the plasma in the region between the two shocks \citep{zhang2005}. While being a good approximation for blastwaves powered by short-lived central engines \citep{ai2021}, this pressure balance model leads in general to energy conservation discrepancies for adiabatic blastwaves. In its stead, a mechanical model assuming a non-uniform pressure in the blastwave has been employed, leading to more physically consistent results regarding the blastwave's evolution \citep{ai2021, uhm2011, beloborodov2006}. An accurate determination of the jump conditions for relativistic shocks is also the basis for the solution of the Riemann problem in relativistic MHD, which is fundamental for the development and testing of Godunov-type numerical schemes \citep{godunov1959} for the solution of the full set of time-dependent RMHD equations \citep{giacomazzo2006}.

In this paper, we expand upon the theory of magnetohydrodynamic relativistic shocks by working in the more general context of resistive relativistic MHD (RRMHD). The jump conditions for relativistic shocks propagating in magnetized media with a finite electrical conductivity are derived in Sect. \ref{s2}, with an additional equation obtained from the covariant Gauss-Ampère Law. The parameter $\alpha_{m}$ that governs the degree to which the shock affects the medium's electromagnetic field is also determined in this section, along with the solutions' two characteristic limits. The jump conditions are solved numerically in the unshocked medium's rest frame, assuming one-dimensional (1D) shock propagation and a transverse electromagnetic field. Approximate analytical solutions to the jump conditions are also derived in the strongly relativistic regime. In Sect. \ref{s3}, we apply the above in numerically solving the Riemann problem in the context of RRMHD. The solution is applied to the study of the ejecta-CBM interaction in its initial phase. The conditions for the formation of a reverse shock are provided, and the blastwave's total pressure and four-velocity are determined by solving the FS and RS jump conditions and demanding the total pressures and flow velocities of the shocked ejecta and shocked CBM to be equal on the contact discontinuity, under the assumption of cold, magnetized and ultra-relativistic ejecta. With the blastwave's four-velocity determined, the FS and RS propagation four-velocities can also be calculated. A calculation of the efficiency of this mechanism for the conversion of the ejecta kinetic energy to thermal energy inside the blastwave is also provided. Our conclusions are presented in Sect. \ref {s4}, along with a discussion of the possible applications of this work in the construction of Riemann solvers for the equations of RRMHD.
\section{The shocked medium}\label{s2}

\subsection{Derivation of the jump conditions}

Following the systematic method proposed by \citet{lich1967}, we derived the covariant jump conditions for a shock propagating with relativistic velocity through a magnetized medium characterized by finite electrical conductivity in flat space-time. The covariant conservation laws describing the fluid and electromagnetic field are:
\begin{equation}
N^{\mu}_{;\mu}=0\, , \label{1}
\end{equation}
\begin{equation}
T^{\mu\nu}_{;\nu}=0\, , \label{2}
\end{equation}
\begin{equation}
\prescript{\ast}{}{F}^{\mu\nu}_{;\mu}=0\, , \label{3}
\end{equation}
\begin{equation}
F^{\mu\nu}_{;\mu}+\dfrac{4\pi}{c}J^{\nu}=0\, , \label{4}
\end{equation}
where $N^{\mu}$ is the particle number four-current density, $T^{\mu\nu}$ is the stress-energy tensor, and $F^{\mu\nu}\, , \prescript{\ast}{}{F}^{\mu\nu}$ are the electromagnetic tensor and its Hodge dual. This system of equations becomes closed by the inclusion of the special relativistic generalization of Ohm's law:
\begin{equation}
    J^{\mu}+J_{\nu}\dfrac{U^{\nu}U^{\mu}}{c^{2}}=\dfrac{\sigma}{c}F^{\mu\nu}U_{\nu}\, ,\label{5}
\end{equation}
with $\sigma$ as the plasma's electrical conductivity. The spatial component ($\mu = i)$ of Eq. \ref{5} gives the expression of Ohm's law in a frame of reference in which the plasma's velocity is $\vec{\beta}$. In the plasma's comoving frame ($\vec{\beta} = 0)$, this expression is:
\begin{equation}
    \vec{J}_{co} = \sigma\vec{E}_{co}\, , \label{6}
\end{equation}
with $\vec{J}_{co}$, $\vec{E}_{co}$ the comoving current density and electric field of the plasma. In a frame in which the plasma has velocity $\vec{\beta}\neq 0$, Ohm's law is expressed as:
\begin{equation}
    \vec{J} + \gamma^{2}\vec{\beta}\left(\vec{\beta}\cdot\vec{J} - J^{0}\right) = \gamma\sigma\left(\vec{E} + \vec{\beta}\cross\vec{B}\right).\,  \label{7}
\end{equation}
The plasma's electrical conductivity $\sigma$ appears as a scalar quantity in Eq. \ref{5} and, as such, it is unaffected by Lorentz boosts between inertial frames of reference. We also assume an equation of state (EoS):
\begin{equation}
    \epsilon=\epsilon(n,P)\, , \label{8}
\end{equation}
where $\epsilon$ is the fluid's total internal energy density and $P$ its thermal pressure. We limit our analysis to flows with transverse electromagnetic fields satisfying $J_{\nu}U^{\nu}=0$. Equation \ref{5} is therefore simplified to
\begin{equation}
    J^{\mu}=\dfrac{\sigma}{c}F^{\mu\nu}U_{\nu}\, \label{9}
\end{equation}
while Eq. 4 becomes:
\begin{equation}
    F^{\mu\nu}_{;\mu}+\dfrac{4\pi\sigma}{c^{2}}F^{\nu\mu}U_{\mu}=0. \label{10}
\end{equation}

The jump conditions are derived by writing Eqs. \ref{1}-\ref{3} and \ref{10} in integral form. The resulting equations are:
\begin{equation}
    [N^{\mu}]S_{\mu}=0\, , \label{11}
\end{equation}
\begin{equation}
    [T^{\mu\nu}]S_{\nu}=0\, , \label{12}
\end{equation}
\begin{equation}
    [\prescript{\ast}{}{F}^{\mu\nu}]S_{\mu}=0\, , \label{13}
\end{equation}
\begin{equation}
    [F^{\mu\nu}]S_{\mu}+\dfrac{4\pi\sigma\mathcal{L}}{c^{2}}\xi(F^{\nu\mu}_{1}U_{1\mu}+F^{\nu\mu}_{2}U_{2\mu})=0\, . \label{14}
\end{equation}
Here $[Q]=Q_{2}-Q_{1}$, while the subscripts $2$ and $1$ denote the boundary value of $Q$ at the shock front in the shocked and unshocked media respectively. Also, $S^{\mu}$ is a spacelike four-vector normal to the shock hypersurface \citep{anile1989}.  Hereafter,  $\sigma$ denotes the electrical conductivity of the plasma in the shock front.

We derived Eq. \ref{14} as follows: we assume that the shock occupies a finite thickness, $\mathcal{L,}$ in Minkowski space, along the shock normal four-vector $S^{\mu}$. Working under this assumption, we approximate the source term in the following way:
\begin{equation}
    \int_{\mathcal{A}}^{\mathcal{B}}\dfrac{4\pi\sigma}{c}F^{\nu\mu}U_{\mu} dl=\dfrac{4\pi\sigma\mathcal{L}}{c}\xi\left(F^{\nu\mu}_{1}U_{1\mu}+F^{\nu\mu}_{2}U_{2\mu}\right)\, , \label{15}
\end{equation}
where $dl$ is the length differential along $S^{\mu}$. $F^{\nu\mu}_{1}U_{1\mu}$, $F^{\nu\mu}_{2}U_{2\mu}$ are the comoving electric field's values at the spacetime points $\mathcal{A}$ and $\mathcal{B,}$ respectively, while $\xi$ is a free parameter, the value of which determines the way $F^{\nu\mu}U_{\mu}$ varies along $S^{\mu}$.

A bounded source term has zero contribution to the jump conditions, when the shock front thickness $\mathcal{L}\rightarrow0$ \citep{leveque2002}. In the particular case of transverse shock propagation in a conductive medium, this bounded source term is the current $\bm{J}_{co}=\dfrac{4\pi\sigma}{c}\bm{E}_{co}$ in the shock front. Its contribution to the jump conditions, particularly the one derived through Ampère's Law is:

\begin{equation}
    \int_{\mathcal{A}}^{\mathcal{B}}\dfrac{4\pi\sigma}{c}F^{\nu\mu}U_{\mu} dl=\dfrac{\mathcal{L}}{\mathcal{L}_{\sigma}}\left\langle F^{\nu\mu}U_{\mu}\right\rangle\, , \label{16}
\end{equation}
 where $\mathcal{L}_{\sigma}$ is the charge relaxation length defined as $\mathcal{L}_{\sigma} = c\tau_{\sigma}$, with $\tau_{\sigma} = \dfrac{1}{4\pi\sigma}$ the charge relaxation time. The source term's contribution to the jump conditions should not be dismissed on the basis that $\mathcal{L}\rightarrow0$. Instead, the degree to which this bounded source term contributes to the jump conditions is determined by the ratio $\dfrac{\mathcal{L}}{\mathcal{L}_{\sigma}}$. If $\mathcal{L}\ll\mathcal{L}_{\sigma}$, then the current has zero contribution to the Ampère's Law jump conditions and the medium's electromagnetic field is unaffected by the shock. For $\mathcal{L}\gg\mathcal{L}_{\sigma}$, the contribution is maximal and the ideal MHD jump conditions are retrieved.

In the following we assume shock propagation along the $\hat{x}$ axis and a transverse electromagnetic field with $\vec{E}_{1,2} = E_{1,2}\hat{y}$ and $\vec{B}_{1,2} = B_{1,2}\hat{z}$. The covariant equations (Eqs. \ref{11}-\ref{14}) provide the following jump conditions in the unshocked plasma's rest frame, where $S^{\mu}=\begin{pmatrix}\varGamma_{s}\beta_{s}&\varGamma_{s}&0&0\end{pmatrix}^{T}$. $\varGamma_{s}$ is the shock Lorentz factor, while $\beta_{s}$ its propagation velocity in units of $c$, as follows:

\begin{equation}
    \gamma_{2}n_{2}(\beta_{s}-\beta_{2})=n_{1}\beta_{s}\, ,\label{17}
\end{equation}
\begin{equation}
\begin{split}
    &\gamma_{2}^{2}(\epsilon_{2}+P_{2})(\beta_{s}-\beta_{2})-P_{2}\beta_{s}+\dfrac{E_{2}^{2}+B_{2}^{2}}{8\pi}\beta_{s}-\dfrac{E_{2}B_{2}}{4\pi}=\\&     \epsilon_{1}\beta_{s}+\dfrac{E_{1}^{2}+B_{1}^{2}}{8\pi}\beta_{s}-\dfrac{E_{1}B_{1}}{4\pi}\, ,\label{18}
\end{split}
\end{equation}
\begin{equation}
\begin{split}
    &P_{2}-\gamma_{2}^{2}(\epsilon_{2}+P_{2})(\beta_{s}-\beta_{2})\beta_{2}+\dfrac{E_{2}^{2}+B_{2}^{2}}{8\pi}-\dfrac{E_{2}B_{2}}{4\pi}\beta_{s}=\\&P_{1}+\dfrac{E_{1}^{2}+B_{1}^{2}}{8\pi}-\dfrac{E_{1}B_{1}}{4\pi}\beta_{s}\, ,\label{19}
\end{split}
\end{equation}
\begin{equation}
    E_{2}-\beta_{s}B_{2}=E_{1}-\beta_{s}B_{1}\, ,\label{20}
\end{equation}
\begin{equation}
    E_{2}\beta_{s}-B_{2}-E_{1}\beta_{s}+B_{1}=\dfrac{4\pi\sigma\xi\mathcal{L}}{c\varGamma_{s}}(\gamma_{2}(E_{2}-\beta_{2}B_{2})+E_{1})\, . \label{21}
\end{equation}
Equations \ref{17}-\ref{21} can be simplified by setting $E_{1}$ and $P_{1}$ equal to zero, conditions usually met in astrophysical propagation media for relativistic shocks.

\begin{figure}[H]
\centering
\resizebox{0.8\columnwidth}{!}{\begin{tikzpicture}
\draw[cyan!25, line width = 12.5] (0.0,-3) -- (0.0,3);
\draw[black, ultra thick] (-0.25,-3) -- (-0.25,3);
\draw[black, ultra thick] (0.25,-3) -- (0.25,3);
\node[draw, circle] at (1.25,-2.5) {\LARGE{1}};
\node[draw, circle] at (-1.25,-2.5) {\LARGE{2}};
\node at (0,-3.5) {\Large{$\tilde{\mathcal{L}}$}};
\node at (1.5,0.5) {\Large{$\vec{\beta}_{s}$}};
\draw [->, very thick] (0.25,0) -- (1.75,0);
\draw[color=black, very thick](3,0) circle (0.25);
\draw[color=black, fill = black!100, very thick](3,0) circle (0.04);
\draw [->, very thick] (3,0.25) -- (3,1.25);
\node at (3,-0.75) {\Large{$\vec{B}_{1}$}};
\node at (2.5,1.0) {\Large{$\vec{E}_{1}$}};
\node at (4.15,0) {\Large{$\vec{\beta}_{1} = 0$}};
\draw[color=black, very thick](-3,0) circle (0.25);
\draw[color=black, fill = black!100, very thick](-3,0) circle (0.04);
\draw [->, very thick] (-3,0.25) -- (-3,1.25);
\draw [->, very thick] (-2.75,0.0) -- (-1.75,0);
\node at (-3,-0.75) {\Large{$\vec{B}_{2}$}};
\node at (-3.5,1.0) {\Large{$\vec{E}_{2}$}};
\node at (-2.0,0.5) {\Large{$\vec{\beta}_{2}$}};
\draw[color=black, very thick](4.5,-2.5) circle (0.25);
\draw[color=black, fill = black!100, very thick](4.5,-2.5) circle (0.04);
\draw [->, very thick] (4.5,-2.25) -- (4.5,-1.25);
\draw [->, very thick] (4.75,-2.5) -- (5.75,-2.5);
\node at (4.5,-3.25) {\Large{$\hat{z}$}};
\node at (4.0,-1.5) {\Large{$\hat{y}$}};
\node at (5.5,-2.0) {\Large{$\hat{x}$}};
\end{tikzpicture}}
\caption{Sketch of the shock front in the general case of a tangential electromagnetic field ($E_{1}\neq0$). $\tilde{\mathcal{L}} = \dfrac{\mathcal{L}}{\varGamma_{s}}$ is the thickness of the shock front in the rest frame of medium 1.}\label{f1}
\end{figure}
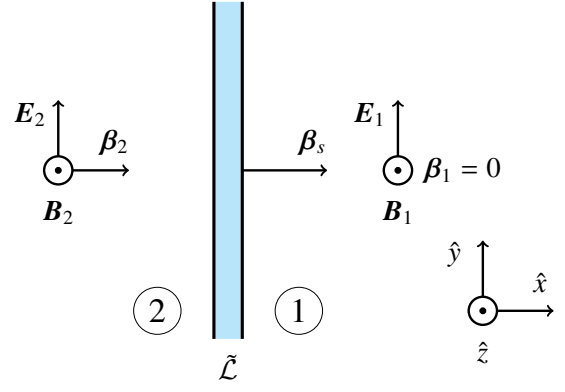

We define the dimensionless parameter $u_{B}$ as:
\begin{equation}
    u_{B}=\dfrac{B_{1}^{2}}{8\pi\epsilon_{1}}\, , \label{22}
\end{equation}
which expresses the magnetic field's energy density in the unshocked fluid, normalized to its rest energy density $\epsilon_{1}=\overline{m}n_{1}c^{2}$. Additionally, the compression ratio $r$ is defined as:
\begin{equation}
    r=\dfrac{\gamma_{2}n_{2}}{n_{1}}=\dfrac{\beta_{s}}{\beta_{s}-\beta_{2}}\, .\label{23}
\end{equation}

We define the dimensionless parameter $\alpha_{m}$ as:
\begin{equation}
    \alpha_{m}=\dfrac{4\pi\xi\sigma\gamma_{2}\varGamma_{s}\beta_{s}\mathcal{L}}{c}\dfrac{\beta_{s}-\beta_{2}}{\beta_{s}}=\dfrac{c\xi\gamma_{2}\varGamma_{s}(\beta_{s}-\beta_{2})\mathcal{L}}{\eta_{m}}\, , \label{24}
\end{equation}
with $\eta_{m}=\dfrac{c^{2}}{4\pi\sigma}$ as the magnetic diffusivity of the plasma in the shock front's volume. By defining $\kappa=\dfrac{\gamma_{2}(\beta_{s}-\beta_{2})}{\tilde{\gamma}(\beta_{s}-\tilde{\beta})}$, with $\tilde{\gamma}\tilde{\beta}$ the four velocity of the plasma inside the shock front, $\alpha_{m}$ becomes:
\begin{equation}
    \alpha_{m}=\kappa\xi\dfrac{c\tilde{\gamma}(\beta_{s}-\tilde{\beta})\mathcal{L}}{\eta_{m}}=\kappa\xi\dfrac{\tilde{\varGamma_{s}}^{2}c\tilde{\beta}_{s}\tilde{\mathcal{L}}}{\eta_{m}}\, , \label{25}
\end{equation}
where $\tilde{\varGamma_{s}}\tilde{\beta}_{s}$ is the shock's four-velocity in the comoving frame of the plasma in the shock front's volume, while $\tilde{\mathcal{L}}$ is the shock front's thickness in that same frame. Also, $\alpha_{m}$ can be expressed more simply as:
\begin{equation}
     \alpha_{m}=\xi\varGamma_{s}\gamma_{2}(\beta_{s}-\beta_{2})\dfrac{\mathcal{L}}{\mathcal{L}_{\sigma}} \, .\label{26}
\end{equation}

Solving Eqs. \ref{20} and \ref{21} for $B_{2}$, $E_{2}$ we derive the following expressions:
\begin{equation}
    B_{2}=\dfrac{1+\alpha_{m}r}{1+\alpha_{m}}B_{1}\, , \label{27}
\end{equation}
\begin{equation}
    E_{2}=\dfrac{\alpha_{m}r}{1+\alpha_{m}}\beta_{2}B_{1}\, . \label{28}
\end{equation}
The comoving electric field, defined as:
\begin{equation}
    \bm{E}_{2}^{co}=\gamma_{2}(\bm{E}_{2}+\bm{\beta}_{2}\cross\bm{B}_{2})\, , \label{29}
\end{equation}
is:
\begin{equation}
    E_{2}^{co}=\gamma_{2}(E_{2}-\beta_{2}B_{2})=-\dfrac{\gamma_{2}\beta_{2}B_{1}}{1+\alpha_{m}}\, . \label{30}
\end{equation} 

We identified two characteristic regimes with respect to the shocked medium's electromagnetic field, for $\alpha_{m}\rightarrow0$ and $\alpha_{m}\rightarrow\infty$, which we call the current-free and ideal MHD regimes, respectively. In the current-free regime, no current layer forms inside the shock front. As a consequence, the propagation medium's electromagnetic field components are unaffected by the shock:
\begin{equation}
    E_{2}=0\, , \label{31}
\end{equation}
\begin{equation}
    B_{2}=B_{1}\, , \label{32}
\end{equation}
while the shocked medium's comoving electric field is:
\begin{equation}
    E_{2}^{co}=-\gamma_{2}\beta_{2}B_{1}\, . \label{33}
\end{equation}
Additionally, Eqs. \ref{17}-\ref{19} in this regime simplify to the jump conditions for shocks in purely hydrodynamic media. For $\alpha_{m}\rightarrow0$ or equivalently $\mathcal{L}\ll\mathcal{L}_{\sigma}$, the propagation medium acts as a hydrodynamic one regardless of its magnetization. In the limit $\alpha_{m}\rightarrow\infty$ or $\mathcal{L}\gg\mathcal{L}_{\sigma}$, the comoving electric field of the shocked medium becomes zero and the ideal MHD jump conditions are retrieved, with $B_{2}$ and $E_{2}$ given by:
\begin{equation}
    B_{2}=\dfrac{\beta_{s}}{\beta_{s}-\beta_{2}}B_{1}\, , \label{34}
\end{equation}
\begin{equation}
    E_{2}=\dfrac{\beta_{2}\beta_{s}}{\beta_{s}-\beta_{2}}B_{1}\, . \label{35}
\end{equation}

The integrated over the shock thickness current density responsible for the change in the plasma's electromagnetic field is:
\begin{equation}
    K=-\dfrac{c}{4\pi}(B_{2}^{s}-B_{1}^{s})\, , \label{36}
\end{equation}
with $B_{1,2}^{s}$ the plasma's magnetic field in the shock frame, given by:
\begin{equation}
    B_{1}^{s}=\varGamma_{s} B_{1}\, , \label{37}
\end{equation}
\begin{equation}
    B_{2}^{s}=\varGamma_{s}(B_{2}-\beta_{s}E_{2})\, . \label{38}
\end{equation}
Using Eqs. \ref{27} and \ref{28} for $B_{2}$ and $E_{2}$, respectively, we arrive at the following expression for $K$:
\begin{equation}
    K=\dfrac{c}{4\pi}\dfrac{\alpha_{m}}{1+\alpha_{m}}(\dfrac{\beta_{s}-\beta_{s2}}{\beta_{s2}})\varGamma_{s} B_{1}\, , \label{39}
\end{equation}
with $\beta_{s2}=\dfrac{\beta_{s}-\beta_{2}}{1-\beta_{s}\beta_{2}}$ the shock propagation velocity in the shocked plasma's frame.

For adequately small values of $\alpha_{m}$, the plasma exits the shock front with a non-zero comoving electric field. There is consequently a non-zero current distribution in the medium exiting the shock front. The methodology presented in this work provides a way of calculating the boundary values of the shocked medium's electromagnetic field components,  as well as its hydrodynamic quantities at the surface through which the fluid exits the shock front.

The solutions presented next were derived by solving Eqs. \ref{17}-\ref{21} for $\{n_{2},P_{2},\epsilon_{2},B_{2},E_{2}\}$ for given density, $n_{1}$, thermal pressure, $P_{1}$, and magnetic field, $B_{1}$, of the propagation medium. The resulting expressions are substituted into the EoS (detailed in the following subsection), which is then solved numerically for $\gamma_{2}$. The rest of the quantities describing the shocked medium are subsequently determined.

\subsection{The Taub-Mathews equation of state}

Assuming shock propagation in the cold interstellar medium, the unshocked gas can be considered cold, in the sense that its thermal pressure is negligible compared to its rest mass energy density. Depending now on the shock's Lorentz factor and the unshocked plasma's magnetic field strength, the shocked fluid can reach relativistic ($\Theta_{2}\gg1$) or non-relativistic ($\Theta_{2}\ll1$) temperatures, where $\Theta_{2}=\dfrac{P_{2}}{\overline{m}n_{2}c^{2}}$ is the dimensionless temperature. The choice of an EoS with a constant adiabatic index, $\hat{\gamma }_{2}$, for the shocked gas can then lead to inaccurate results, as its value can be anywhere between $\hat{\gamma}_{2}=\dfrac{5}{3}$ and $\hat{\gamma}_{2}=\dfrac{4}{3}$, depending on the strength of the shock. This can be avoided by assuming an EoS with a variable adiabatic index, $\hat{\gamma}_{2}=\hat{\gamma}_{2}(\Theta_{2})$.

The EoS used in this work is the Taub-Mathews EoS, which provides the following expression for the specific enthalpy:
\begin{equation}
    h_{2}=\dfrac{5}{2}\Theta_{2}+\sqrt{\dfrac{9}{4}\Theta_{2}^{2}+1}\, . \label{40}
\end{equation}
This expression is derived by demanding that the equality in Taub's fundamental inequality,
\begin{equation}
    (h_{2}-\Theta_{2})(h_{2}-4\Theta_{2})\geq1\, , \label{41}
\end{equation}
is satisfied \citep{mignone2007}.
The variable adiabatic index as a function of the temperature $\Theta_{2}$ is:
\begin{equation}
    \hat{\gamma}_{2}=\dfrac{5\Theta_{2}+\sqrt{9\Theta_{2}^{2}+4}-2}{3\Theta_{2}+\sqrt{9\Theta_{2}^{2}+4}-2}\, , \label{42}
\end{equation}
which for $\Theta_{2}\ll1$ and $\Theta_{2}\gg1$ takes the values $\hat{\gamma}_{2}=\dfrac{5}{3}$ and $\hat{\gamma}_{2}=\dfrac{4}{3}$ respectively. Consequently, the shocked fluid's total internal energy density is related to its thermal pressure through:
\begin{equation}
    \epsilon_{2}=\dfrac{3}{2}P_{2}+\sqrt{\dfrac{9}{4}P_{2}^{2}+\overline{m}^{2}n_{2}^{2}c^{4}}\, . \label{43}
\end{equation}

\subsection{Numerical solutions}\label{sub1}

In the case of a strongly relativistic shock ($\varGamma_{s}\sim100$), for $\alpha_{m}\ll1$, the shocked plasma's Lorentz factor, compression ratio, and thermal pressure coincide with the results of \citet{bm1976} for ultra-relativistic shocks in hydrodynamic media regardless of the actual value of $u_{B}$ for the unshocked plasma, while the plasma's electric and magnetic fields are unaffected by the shock. The post-shock Lorentz factor, compression ratio, and thermal pressure become less significant as the value of $\alpha_{m}$ increases. 

\begin{figure}[H]
 \centering
 \includegraphics[width=1\linewidth]{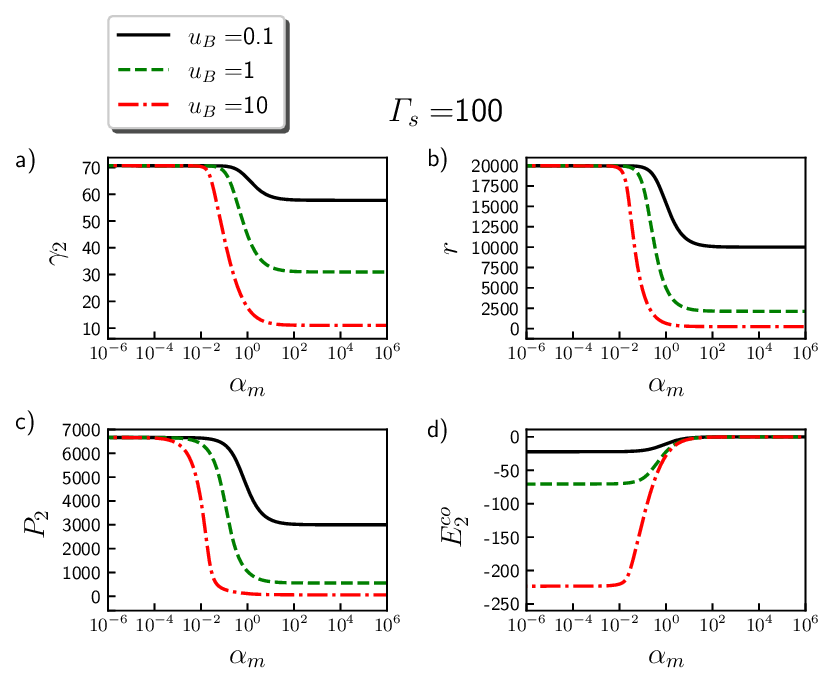}
 \caption{Shocked plasma quantities with respect to $\alpha_{m}$ for a shock propagating with $\varGamma_{s} = 100$: a) shocked plasma Lorentz factor $\gamma_{2}$, b) compression ratio $r$, c) thermal pressure $P_{2}$ in units of $\epsilon_{1}$, and d) shocked medium comoving electric field in units of $\sqrt{8\pi\epsilon_{1}}$ with respect to $\alpha_{m}$.}\label{f2}
\end{figure}

\begin{figure}[H]
 \centering
 \includegraphics[width=1\linewidth]{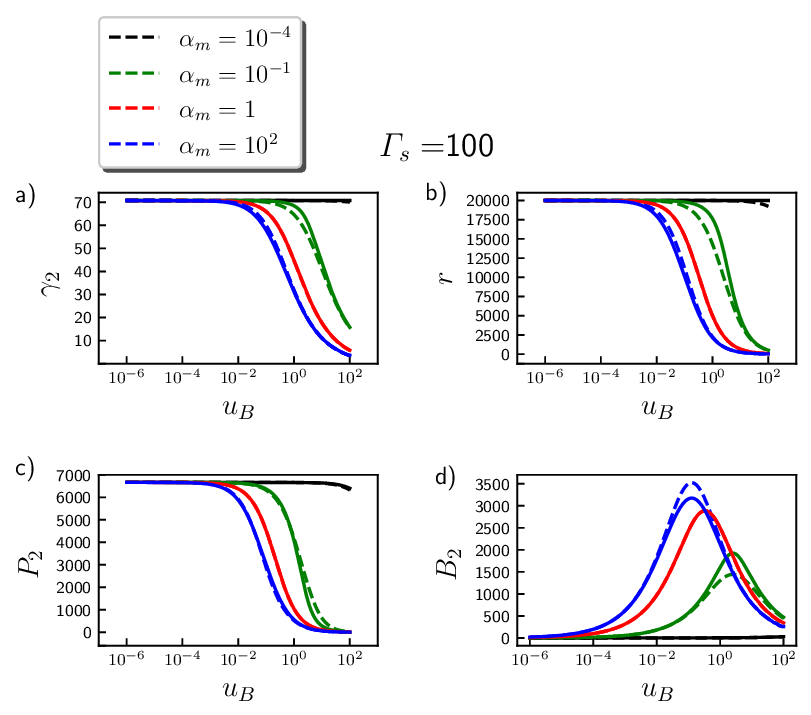}
 
 
 \caption{Comparison between the numerical solutions to the jump conditions and the respective approximate analytical expressions. The solid lines correspond to the numerical solutions, while the dashed to the analytical approximations.} \label{f4} 
\end{figure}

For large values of this parameter, all post-shock quantities exhibit the behavior predicted by the ideal MHD jump conditions, which can be derived from Eqs. \ref{17}-\ref{21} by using the expression given by Eq. \ref{30} for the shocked plasma's comoving electric field in Eq. \ref{21}, for $\alpha_{m}\rightarrow\infty$. This is clearly seen in Fig. \ref{f4}, where we can observe that the solutions to Eqs. \ref{17}-\ref{21} for $\alpha_{m}=10^{2}$ coincide with the ideal MHD solutions. The electrical conductivity $\sigma$ in the shock front increases with $\alpha_{m}$, with its increase becoming less steep for larger magnetic energy densities of the unshocked plasma, in the high $\alpha_{m}$ regime. As the value of $\alpha_{m}$ gets larger, the shocked plasma's comoving electric field asymptotically approaches zero.

The behavior exhibited by the shocked electric and magnetic fields can be better understood by defining an effective magnetization for the propagation medium as:
\begin{equation}
    u_{B}^{eff}=\left(\dfrac{1+\alpha_{m}r}{r+\alpha_{m}r}\right)^{2}u_{B}\, .\label{44}
\end{equation}
This effective magnetization's value ranges from $u_{B}^{eff}=\dfrac{u_{B}}{r^{2}}$ to $u_{B}$, depending on the value of $\alpha_{m}$. For intermediate $\alpha_{m}$ values and significant magnetization $u_{B}\geq0.1$, the shocked magnetic and electric field display a maximum, due to the effective magnetization of the propagation medium being smaller than its actual magnetization. It must be noted though that the behavior of the shocked fluid's electromagnetic field for a given effective magnetization $u_{B}^{eff}$ is not equivalent to the case of a shock with $\alpha_{m}\rightarrow\infty$ propagation in a medium with magnetization $u_{B}=u_{B}^{eff}$. In the latter case, the shocked comoving electric field would be equal to zero.

\begin{figure}[H]
 \centering
  
  \includegraphics[width=1\linewidth]{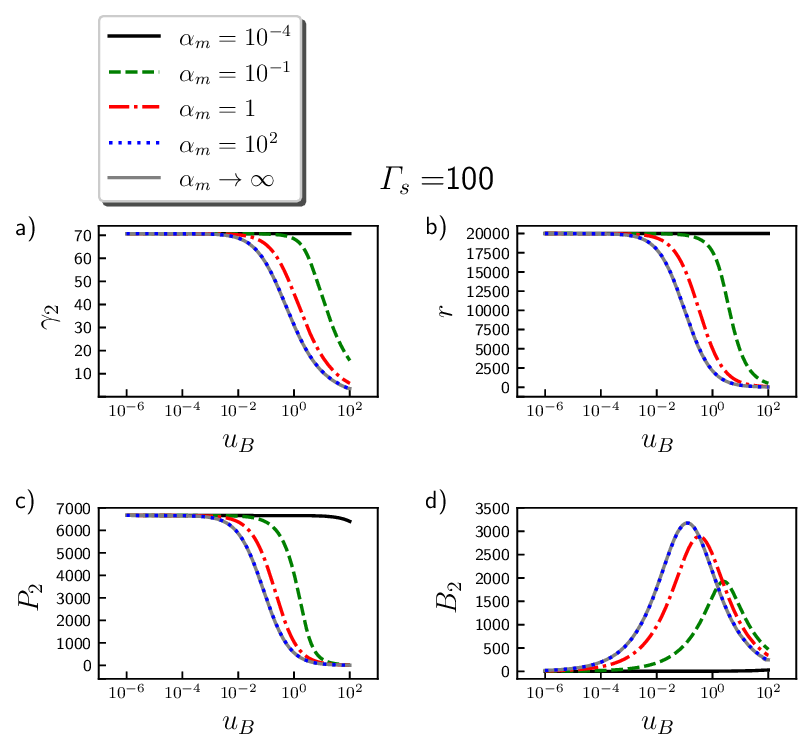}
   \caption{Shocked plasma quantities with respect to $u_{B}$ for a shock propagating with $\varGamma_{s} = 100$: a) shocked plasma Lorentz factor $\gamma_{2}$, b) compression ratio $r$, c) thermal pressure $P_{2}$, and d) magnetic field $B_{2}$ with respect to the propagation medium's magnetization $u_{B}$, for various $\alpha_{m}$ values.} \label{f3}
\end{figure}

The effects of $\alpha_{m}$ on the post-shock plasma were also studied for the case of mildly relativistic shock propagation ($\varGamma_{s}=5$) in a plasma with the same characteristics as the propagation medium of Sect. \ref{sub1}. The effects of $\alpha_{m}$ on the post-shock medium are qualitatively the same as in the case of strongly relativistic shock propagation. As $\alpha_{m}$ increases, the effective magnetic energy density $u_{B}^{eff}$ increases, asymptotically approaching  the actual magnetic energy density $u_{B}$ of the propagation medium, while the the shocked plasma's comoving electric field approaches zero, satisfying the ideal MHD condition for $\alpha_{m}\rightarrow\infty$. For both strongly and mildly relativistic shock propagation, the plasma's electromagnetic field is practically unaffected for small $\alpha_{m}$ values.

\subsection{Analytical expressions in the ultra-relativistic regime}\label{sub2}

Working under the assumptions of a strongly relativistic shock propagating in a cold ($P_{1}=0$) magnetized medium, ultra-relativistic motion ($\gamma_{2}\gg1$) of the shocked plasma in the unshocked medium's frame, and of an adiabatic index $\hat{\gamma}_{2}=\dfrac{4}{3}$ for the shocked plasma, \citet{zhang2005} derived an analytical expression for the shocked plasma's relative four-velocity to that of the shock, determining it as a function of only the propagation medium's magnetization. Approximate analytical expressions in the ultra-relativistic regime were also derived by \citet{lyutikov2002} in the low and high magnetization limits.

We adopted the same assumptions and derive approximate analytical expressions for all post-shock quantities in the unshocked medium's frame. We assume a constant adiabatic index equal to $\dfrac{4}{3}$ for the shocked plasma and expand the EoS to terms up to $\order{\varGamma_{s}^{-2}}$, $\order{\gamma_{-2}^{2}}$, obtaining a significantly simplified expression, which is then solved for $\gamma_{2}^{2}$. The method for the derivation of the approximate analytical expressions for all post-shock quantities is explained in greater detail in Appendix \ref{appa}, along with the relevant expressions. 

For low $\alpha_{m}$ values, the analytical expressions slightly underestimate the values of $\gamma_{2}$, $r$, and $P_{2}$, with the most significant deviation found in the compression ratio, $r$. The greatest deviation from the numerical solutions is observed for $\alpha_{m}=10^{-1}$, being most prominent for the shocked plasma's magnetic field $B_{2}$, the value of which is underestimated more than those of $\gamma_{2}$, $r$, and $P_{2}$ by its analytical expression. However, this non-negligible deviation of the analytical expressions from the numerical solutions observed in certain cases is offset by their simplicity. It should also be noted that in the limit $\alpha_{m}\rightarrow\infty$, expressions \ref{51}-\ref{54} reduce to the analytical expressions for ideal MHD shocks, which are presented in Appendix \ref{appb}.

\section{Application of the Riemann problem  to gamma-ray bursts} \label{s3}

The Riemann problem consists of two fluids generally characterized by different velocities, densities, pressures and, in the case of magnetized plasma, electromagnetic fields, separated by a surface of tangential discontinuity. Through the interaction of the two fluids a wave system emerges, consisting of a contact discontinuity (CD) and any possible combination of a left and right-propagating, in the CD frame, rarefaction wave or shock front \citep{toro1997}. In the case of two propagating shock fronts, which is of interest here, the RS and FS shock conditions are solved simultaneously. The demand for equal total pressure in regions 2 and 3 (shown in Fig. \ref{f5}) determines the kinematic and dynamical conditions inside the blastwave, the volume of fluid contained between the two shocks \citep{sari1995}, assuming a constant Lorentz factor for the fluid in the blastwave \citep{kobayashi2000}. We applied this method to determine the dependence of these conditions inside the blastwave, as well as of the RS and FS propagation four-velocities, on the value of $\alpha_{m}$.

\subsection{Reverse shock formation}

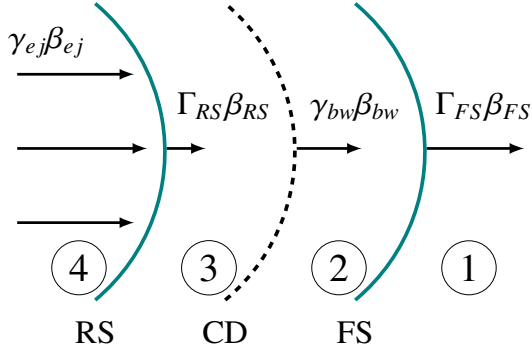
\begin{figure}[H]
\centering
\resizebox{0.8\columnwidth}{!}{
\begin{tikzpicture}
  \coordinate (I)  at (1.9,0);     
  \path [name path=arc1, draw=none](-1.9,-1.5) 
     arc[start angle=-90, end angle=90,radius=1.5];
  \path [name path=arc2, draw=none](1.99,-1.5)   
     arc[start angle=90, end angle=270,radius=3];
  \path [name path=rect, draw=none](-0.9,-0.9) rectangle (0.9,0.9);

  \path [name intersections={of = arc1 and rect}];
  \coordinate (A)  at (intersection-1);
  \coordinate (B)  at (intersection-2);

  \path [name intersections={of = arc2 and rect}];
  \coordinate (C)  at (intersection-1);
  \coordinate (D)  at (intersection-2);

  \pgfmathanglebetweenpoints{\pgfpointanchor{I}{center}}{%
    \pgfpointanchor{D}{center}} 
  \let\tmpan\pgfmathresult 
  \draw [teal, ultra thick] (-1.5,-5.5) arc[start angle=-50, end angle=50,radius=3];
  \draw [black, ultra  thick, dashed] (0.5,-5.5) arc[start angle=-50, end angle=50,radius=3];
  \draw [teal, ultra thick] (2.5,-5.5) arc[start angle=-50, end angle=50,radius=3];

  \begin{scope}[>=latex]
    \draw [->, very thick] (3.6,-3.15) -- (5.1,-3.15);
    \draw [->, very thick] (1.6,-3.15) -- (2.6,-3.15);
    \draw [->, very thick] (-0.4,-3.15) -- (0.2,-3.15);
    \draw [->, very thick] (-2.7,-3.15) -- (-0.7,-3.15);
    \draw [->, very thick] (-2.7,-2.0) -- (-0.9,-2.0);
    \draw [->, very thick] (-2.7,-4.3) -- (-0.9,-4.3);

  \end{scope}
  \node at (4.5,-2.5) {\Large{$\Gamma_{FS}\beta_{FS}$}};
  \node at (2.5,-2.5) {\Large{$\gamma_{bw}\beta_{bw}$}};
  \node at (0.5,-2.5) {\Large{$\Gamma_{RS}\beta_{RS}$}};
  \node at (-2.25,-1.5) {\Large{$\gamma_{ej}\beta_{ej}$}};
  \node at (2.5,-6.0) {\Large{FS}};
  \node at (0.5,-6.0) {\Large{CD}};
  \node at (-1.5,-6.0) {\Large{RS}};
  \node[draw, circle] at (4.25,-5.0) {\LARGE{1}};
  \node[draw, circle] at (2.25,-5.0) {\LARGE{2}};
  \node[draw, circle] at (0.25,-5.0) {\LARGE{3}};
  \node[draw, circle] at (-1.75,-5.0) {\LARGE{4}};
\end{tikzpicture}}

 \caption{Illustration of the ejecta-unshocked CBM interaction in the latter's rest frame. The forward and reverse shocks travel towards the same direction in the unshocked CBM frame with four-velocity lower than that of the ejecta. The contact discontinuity, represented by the dashed line, is the surface on which the total pressures of the shocked CBM and shocked ejecta are equal. The arrows represent the four-velocity of the forward shock, shocked material, reverse shock, and ejecta.} \label{f5}
 \end{figure}

When a GRB occurs, magnetized plasma expelled at ultra-relativistic velocity from the central engine interacts with the cold CBM, gradually decelerating through this interaction, while a forward shock propagating through the CBM is always formed, due to the CBM's negligible thermal and magnetic pressures. In most cases, a reverse shock propagating in the ejecta is also formed. The full picture of the ejecta-CBM interaction is generally similar to the one presented in Fig. \ref{f5}, consisting of a forward and a reverse shock propagating with four velocities of  $\varGamma_{FS}\beta_{FS}$ and $\varGamma_{RS}\beta_{RS}$ (respectively) through the CBM and of a region enclosed by the two shock fronts containing the shocked CBM and ejecta. The surface represented by the dashed line is the contact discontinuity, the surface on which the shocked CBM and ejecta are at a dynamical equilibrium, that is, their total pressures are equal. The phenomenon's driving mechanism is the energy of the ejecta, which act as a piston, compressing the ambient medium surrounding the central engine.

The existence of the RS cannot be taken as a given as in the case of the FS. Generally, in order for shock propagation to occur in a magnetized medium two conditions must be met. Firstly, the unshocked ejecta's velocity must exceed the local fast magnetosonic velocity. Secondly, the shocked plasma's total pressure must be larger than that of the propagation medium's \citep{zhang2005}. Concerning the first condition, for an electromagnetic field tangential to the plasma's flow velocity, the fast magnetosonic Lorentz factor for magnetized ejecta with an adiabatic index of $\hat{\gamma}_{ej}$ given by Eq. \ref{42} is:
\begin{equation}
    \gamma_{ej}^{fast}=\sqrt{\dfrac{6h_{ej}(1+2u_{ej}^{B})}{3h_{ej}^{2}+h_{ej}-(h_{ej}-1)\sqrt{9h_{ej}^{2}+16}+2}}\, , \label{45}
\end{equation}
with $h_{ej}$ as the unshocked ejecta specific enthalpy and $2u_{ej}^{B}=\dfrac{2P_{ej}^{m}}{\overline{m}n_{ej}h_{ej}c^{2}}$ as their magnetization. Requiring that the ejecta flow Lorentz factor exceed $\gamma_{ej}^{fast}$, we find the following upper limit for their magnetization:
\begin{equation}
    2u_{ej}^{B}\leq\gamma_{ej}^{2}\dfrac{3h_{ej}^{2}+h_{ej}-(h_{ej}-1)\sqrt{9h_{ej}^{2}+16}}{6h_{ej}}\, . \label{46}
\end{equation}
For cold ejecta with $h_{ej}\approx1$ the following simplified condition is determined:
\begin{equation}
    2u_{ej}^{B}\leq\gamma_{ej}^{2}\dfrac{3h_{ej}^{2}-h_{ej}+13}{15h_{ej}}\, . \label{47}
\end{equation}

\vspace{-1.625cm}

The second condition constrains the ejecta magnetization to significantly lower values. By assuming that $\varGamma_{FS}\simeq\gamma_{ej}$ as well as $u_{CBM}^{B}=0$, the pressure in the shocked region, generated by the FS, is $P_{FS}=\dfrac{2\gamma_{ej}^{2}}{3}n_{CBM}c^{2}$. The ejecta magnetization upper limit is then given by:
\begin{equation}
   2u_{ej}^{B}\leq\dfrac{4\gamma_{ej}^{2}}{3h_{ej}}\dfrac{n_{CBM}}{n_{ej}}-\dfrac{5h_{ej}-\sqrt{9h_{ej}^{2}+16}}{4h_{ej}}\, . \label{48}
\end{equation}
For cold ejecta $h_{ej}\approx1$, this relation is reduced to:
\begin{equation}
    2u_{ej}^{B}\leq\dfrac{4\gamma_{ej}^{2}}{3h_{ej}}\dfrac{n_{CBM}}{n_{ej}}-\dfrac{4(h_{ej}-1)}{5h_{ej}}\, . \label{49}
\end{equation}

The interaction however between the CBM and the ejecta takes place after the expelled plasma has been accelerated thermally and magnetically and so the value of $h_{ej}$ is close to $1$. The conditions for reverse shock formation in phases after the initial collision, when the ejecta have been decelerated through their interaction with the CBM,  have been studied in detail in works such as \citet{zhang2005} and \citet{giannios2008}.

\begin{figure}[H]
 \centering
  \includegraphics[width=1.0\linewidth]{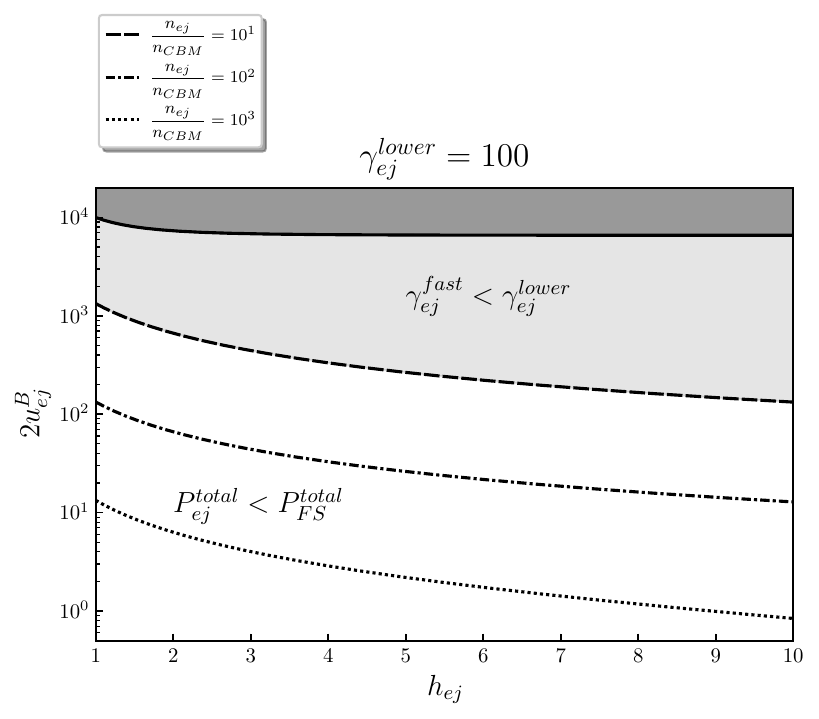}
 
   \caption{Demanding a lower value of $100$ for $\gamma_{ej}$ \citep{zhang2005}, Eq. \ref{46} provides an upper limit for $u_{ej}^{B}$. The requirement that the total pressure generated by the FS be stronger than the total pressure of the ejecta significantly lowers the value of this upper limit. Depending on the ratio between the unshocked CBM and ejecta baryon number densities, this upper limit can realistically range from $\order{10^{2}}$ to $\order{10^{1}}$.}\label{f6}
\end{figure}

Figure \ref{f6} shows the ejecta magnetization values satisfying the conditions for which a reverse shock can form, as a function of the ejecta specific enthalpy. The light gray area represents the space of specific enthalpy-ejecta magnetization values for which only the first condition is satisfied. The white area below each line corresponding to a different value of $\dfrac{n_{ej}}{n_{CBM}}$ is the set of values of these two parameters for which both conditions are satisfied. For example, for $\dfrac{n_{ej}}{n_{CBM}} = 10^{2}$, both conditions are satisfied and a reverse shock can form for ejecta magnetization values below the dot-dashed line.

\subsection{Determining the FS, RS, and blastwave four-velocities}
Under the assumption that all conditions leading to reverse shock formation are met, the interaction between the ejecta and the CBM is represented by Fig. \ref{f5}. Region 2 (shocked CBM) and region 3 (shocked ejecta) are separated by a surface of discontinuity, upon  which the flow velocities and total pressure of the plasma in each region are equal. By solving the jump conditions for both shocks and assuming that regions 2 and 3 (shown in Fig. \ref{f5}) are uniform, we may determine the common pressure and four-velocities of the shocked CBM and ejecta, as well as the four-velocities of both the FS and the RS at the onset of the interaction between the CBM and the ejecta, for a given initial ejecta Lorentz factor, $\gamma_{ej}$. In our solutions, we chose a typical value of $\gamma_{ej}=500$ for the unshocked ejecta.

\begin{figure}[H]
\centering
\subfloat{
\includegraphics[width=0.4\textwidth]{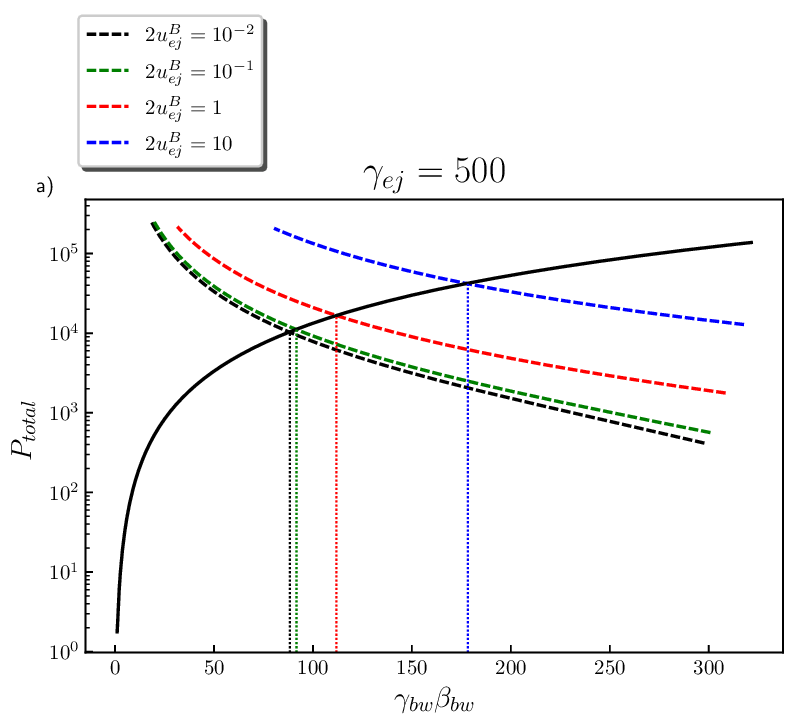}
\label{fig:subfig1}}
\qquad
\subfloat{
\includegraphics[width=0.4\textwidth]{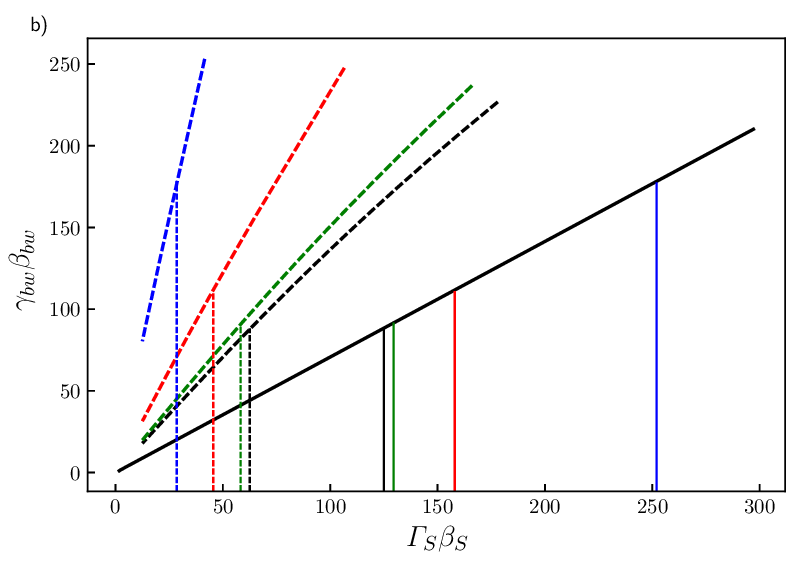}
\label{fig:subfig2}}
\caption{ FS and RS four-velocities in the ideal MHD limit $\alpha_{m}\rightarrow\infty$, for unshocked ejecta with $\gamma_{ej}=500$, $h_{ej}=1.001$, $\dfrac{n_{ej}}{n_{CBM}}=1000$. Higher ejecta magnetizations lead to greater four-velocities for the plasma inside the blastwave, as well as for the forward shock. For $2u_{ej}^{B}=10$ the shocked ejecta total pressure is stronger than the thermal pressure generated by the FS for all possible FS propagation velocities. The dashed lines denote RS quantities.}\label{f7}
\end{figure}

The points at which the FS thermal pressure curve intersects with RS total pressure curves determine the pressure and four-velocity at the contact discontinuity (Fig. \ref{fig:subfig1}). The shocked plasma's four-velocity at the contact discontinuity is determined by plotting the shocked CBM and shocked ejecta four-velocities over the shock propagation four-velocities, as seen in Fig. \ref{fig:subfig2}.

\begin{figure}[H]
\centering
\subfloat{
\includegraphics[width=0.4\textwidth]{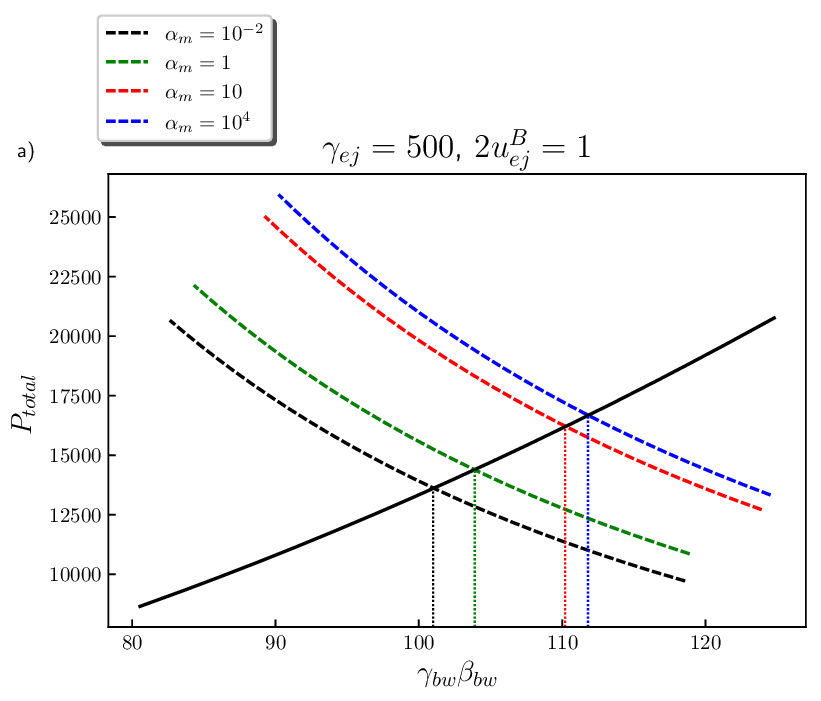}
\label{fig:subfig3}}
\qquad
\subfloat{
\includegraphics[width=0.4\textwidth]{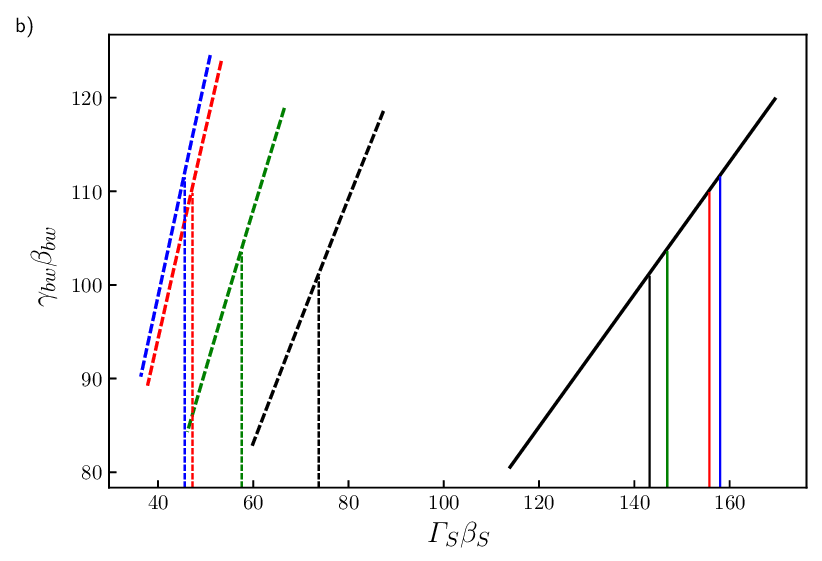}
\label{fig:subfig4}}
\caption{Total pressure and four-velocity of the blastwave are significantly affected by the value of the RS, $\alpha_{m}$, for a given ejecta magnetization. Higher $\alpha_{m}$ values require stronger forward and reverse shocks. The dashed lines once again correspond to the RS, while the solids to the FS. The unshocked ejecta are characterized by $h_{ej}=1.001$, $\dfrac{n_{ej}}{n_{CBM}}=1000$.}\label{f8}
\end{figure}

Depending on the value of $\alpha_{m}$, the blastwave's four-velocity displays a significant variation, with plasma in the region between the RS and FS becoming faster as $\alpha_{m}$ becomes larger, as a result of the magnetic pressure generated by the RS becoming stronger with increasing $\alpha_{m}$ values. The classical ideal MHD behavior is retrieved for high enough $\alpha_{m}$ values ($\alpha_{m}=10^{4}$ in the solutions to the FS-RS jump conditions, presented in Fig. \ref{f8}).


The evolution of the blast wave at later times can then be determined. As a number of works have shown \citep{beloborodov2006, uhm2011, ai2021}, the customary pressure balance model for determining the pressure and velocity of the plasma inside the blast wave leads to discrepancies with regards to total energy conservation when the ejecta are powered by long-lived central engines \citep{ai2021}. The aforementioned works employ a 'mechanical' model instead, which provides more accurate results at large distances from the central engine. This kind of analysis is beyond the scope of this work. We base our modelling on the assumption of a short-lived central engine in order to investigate the effects of the $\alpha_{m}$ value characterizing the RS on the conditions inside the blast wave for a given ejecta magnetization.

The efficiency, $\chi,$ of the conversion of the total ejecta energy to blastwave internal energy through the reverse shock is calculated as:

\begin{equation}
    \chi = \dfrac{\epsilon_{3}}{T^{00}_{ej}}\dfrac{M_{ej}}{M_{3}}\, ,
\end{equation}
with $M_{ej}$ as the mass of the ejecta and $M_{3}$ the mass of the matter enclosed in area 3 at a given time $t$, calculated as $M_{ej} = m n_{ej} c\beta_{ej} t$, $M_{3} = m n_{3} c (\beta_{CD} - \beta_{RS}) t$. The values of $\chi$ for the four solutions presented above are given in Table 1. Smaller values of $\alpha_{m}$ can then lead to better efficiencies in the
conversion of the total ejecta energy into blastwave thermal energy.
\begin{table}[H]
    \centering
    \caption*{\textbf{Table 1.} RS efficiencies} \label{tab:title} 
    \begin{tabular}{|c|c|}
    \specialrule{1.5pt}{0pt}{0pt}
    $\alpha_{m}$  & $\chi$ \\
    \specialrule{1pt}{0pt}{0pt}
        $10^{-2}$ & $29.6784$ \\
        $10^{-1}$ & $24.3928$ \\
        $1$ & $9.5946$ \\
        $10$ & $7.5291$ \\
        $10^{4}$ & $7.5115$ \\
    \specialrule{1.5pt}{0pt}{0pt}
    \end{tabular}
\end{table}

\section{Conclusions}\label{s4}

The theory of relativistic shock propagation in magnetized media was expanded to the case of propagation media characterized by a finite electrical conductivity. We assumed a finite shock thickness, $\mathcal{L,}$ and supplemented the covariant equations expressing the jump conditions for relativistic shocks with an additional equation derived through the covariant Gauss-Ampère Law. The $3+1$ decomposition of these covariant equations showed that the boundary properties of the shocked medium at the shock front are determined by the value of a dimensionless parameter $\alpha_{m}$ which has a linear dependence on the ratio $\dfrac{\mathcal{L}}{\mathcal{L}_{\sigma}}$. We identified two characteristic regimes for the shocked medium's electromagnetic field corresponding to $\mathcal{L}\ll\mathcal{L}_{\sigma}$ and $\mathcal{L}\gg\mathcal{L}_{\sigma}$. The first of these is the current free regime, in which the propagation medium's electromagnetic field is not affected by the shock, while its hydrodynamic quantities are given by the jump conditions for shocks in non-magnetized media. In the second regime, the jump conditions are reduced to those of ideal MHD.

The dependence of the shocked medium's properties on the ratio $\dfrac{\mathcal{L}}{\mathcal{L}_{\sigma}}$ was investigated in depth through the numerical solution of the jump conditions \ref{17}-\ref{21}. In order to achieve high accuracy in our results we used a non-polytropic EoS, namely, the Taub-Mathews EoS. These numerical solutions allowed us to determine the shocked medium's properties in the full range of values of $\alpha_{m}$.

We also examined the conditions for the formation of a reverse shock during the ejecta-CBM interaction, assuming the variable adiabatic index Taub-Mathews EoS for the ejecta, and determined the dependence of the ejecta magnetization's upper limit on their specific enthalpy. We found that this upper limit decreases as the ejecta's thermal pressure increases, a corollary of the reverse shock formation condition $P_{ej}^{total}>P_{FS}$. For realistic values of the ejecta's specific enthalpy ($h_{ej}\approx1$), their magnetization's upper limit is given by Eq. \ref{49}. 

In addition, assuming the existence of a reverse shock, we solved the ideal FS and RS jump conditions for cold $(h_{ej}=1.001)$ unshocked ejecta with a Lorentz factor $\gamma_{ej}=500$ for various magnetization values, up to $2u_{ej}^{B}=10$. By requiring that the FS and RS generated total pressures be equal inside the blastwave, we determined its four-velocity and consequently the FS and RS propagation four-velocities and their dependence on the ejecta magnetization, in the initial phase of the ejecta-CBM interaction. We found that higher ejecta magnetizations demand that both the FS and RS be stronger in their respective propagation media frames (the CBM for the FS and the unshocked ejecta for the RS). As a consequence, the FS and RS must be faster and slower respectively in the CBM frame as the ejecta magnetization becomes higher. Also, the four-velocity of the plasma in the volume between the two shocks increases with the ejecta magnetization. For ejecta with $\gamma_{ej}=500$, $h_{ej}=1.001$, and $2u_{B}^{ej}=1$ we determined the dependence of the blastwave's and shocks' four-velocities on the RS $\alpha_{m}$. Larger values of $\alpha_{m}$ lead to stronger RS magnetic and total pressures generated by the RS. This in turn leads to higher blastwave and shock four-velocities. For large $\alpha_{m}$ values ($\alpha_{m}=10^{4}$), we retrieve the same conditions on the contact discontinuity as in the case of the ideal MHD jump conditions. One result of great astrophysical significance is that for small $\alpha_{m}$ values, the predicted efficiency of the conversion of the ejecta energy into thermal energy of the blastwave is much larger than in the ideal MHD regime.

Lastly, the potential application of the theory and results presented here in the development of conservative numerical schemes must be stated. Godunov-type conservative numerical schemes relying on the exact or approximate solution of the jump conditions at cell interfaces via the use of Riemann solvers have found extensive applications in modern-day relativistic astrophysics. What our results demonstrate is that the conductivity of the plasma caught inside the shock front, or at the cell interface in the case of a fluid moving with velocity larger than the local characteristic velocities in the respective cell's volume, determines the significance of the current layer formed at the surface of discontinuity. In this way the electrical conductivity greatly impacts the properties of the post-shock medium. Since the accurate solution of the jump conditions is central to the implementation of such numerical methods, the theory developed in this work can assist in the construction of robust and accurate conservative schemes for RRMHD.

\begin{acknowledgements}
      This work was supported in full by the State Scholarships Foundation (IKY) scholarship program from the proceeds of the "Nic. D. Chrysovergis" bequest.
\end{acknowledgements}

%
%

\begin{appendix} \label{app}

\section{Approximate analytical expressions}\label{appa}
Assuming an adiabatic index equal to $\dfrac{4}{3}$ for the shocked medium, as well as $\varGamma_{s}\gg1,$ we derive the following expression for $\gamma_{2}$:
\begin{equation}
   \begin{split}
    \gamma_{2}^{2}=&\dfrac{(\varGamma_{s}^{2}+9u_{B}+2)\alpha_{m}^{2}+(2\varGamma_{s}^{2}+3u_{B}+4)\alpha_{m}+\varGamma_{s}^{2}+2}{4(4u_{B}+1)\alpha_{m}^{2}+8(u_{B}+1)a_{m}+4}+\\&\left(\dfrac{((\varGamma_{s}^{2}+9u_{B}+2)\alpha_{m}^{2}+(2\varGamma_{s}^{2}+3u_{B}+4)\alpha_{m}+\varGamma_{s}^{2}+2)^{2}}{(4(4u_{B}+1)\alpha_{m}^{2}+8(u_{B}+1)a_{m}+4)^{2}}-\right.\\&\dfrac{((4u_{B}+1)\alpha_{m}^{2}+2(u_{B}+1)\alpha_{m}+1)((8\varGamma_{s}^{2}+10u_{B}+1)\alpha_{m}^{2}}{(4(4u_{B}+1)\alpha_{m}^{2}+8(u_{B}+1)a_{m}+4)^{2}}+\\&\left.\dfrac{(16\varGamma_{s}^{2}+-4u_{B}+2)\alpha_{m}+8\varGamma_{s}^{2}+1)}{(4(4u_{B}+1)\alpha_{m}^{2}+8(u_{B}+1)a_{m}+4)^{2}}\right)^{\dfrac{1}{2}}\, ,\label{50}
    \end{split}
\end{equation}
This expression can be further simplified by keeping only the terms of $\order{\varGamma_{s}^{2}}$:
\begin{equation}
    \gamma_{2}^{2}=\dfrac{(\alpha_{m}+1)^{2}}{2(4u_{B}+1)\alpha_{m}^{2}+4(u_{B}+1)\alpha_{m}+2}\varGamma_{s}^{2}\, . \label{51}
\end{equation}
Similarly, we substitute $\gamma_{2}$ with the expression given by Eq. \ref{50} in all other quantities describing the shocked plasma and keep the highest order terms with respect to $\varGamma_{s}^{2}$. The following expressions are derived:
\begin{equation}
    r=\dfrac{2(\alpha_{m}+1)^{2}}{(8u_{B}+1)\alpha_{m}^{2}+2(2u_{B}+1)\alpha_{m}+1}\varGamma_{s}^{2}\, , \label{52}
\end{equation}
\begin{equation}
    \begin{split}
    \dfrac{P_{2}}{\epsilon_{1}}= 2(\alpha_{m}+1)^{2}\varGamma_{s}^{2}\dfrac{\mathcal{N}}{\mathcal{D}}\, , \label{53}
    \end{split}
\end{equation}
with
\begin{equation*}
    \begin{split}
    \mathcal{N} = &(16u_{B}^{2}+6u_{B}+1)\alpha_{m}^{4}+4(2u_{B}^{2}+4u_{B}+1)\alpha_{m}^{3} + \\&  2(7u_{B}+3)\alpha_{m}^{2}+4(u_{B}+1)\alpha_{m}+1\, ,
    \end{split}
\end{equation*}
\begin{equation*}
    \begin{split}
    \mathcal{D} = &((8u_{B}+1)\alpha_{m}^{2}+(4u_{B}+2)\alpha_{m}+1)^{2}\times \\&((8u_{B}+3)\alpha_{m}^{2}+(4u_{B}+6)\alpha_{m}+3)\, .
    \end{split}
\end{equation*}
The expressions for the magnetic and electric field of the shocked plasma are:
\begin{equation}
    \dfrac{B_{2}}{\sqrt{8\pi\epsilon_{1}}}=\dfrac{2\alpha_{m}(\alpha_{m}+1)\varGamma_{s}^{2}+1}{(8u_{B}+1)\alpha_{m}^{2}+2(2u_{B}+1)\alpha_{m}+1}\sqrt{u_{B}}\, , \label{54}
\end{equation}
\begin{equation}
    \dfrac{E_{2}}{\sqrt{8\pi\epsilon_{1}}}=\dfrac{2\alpha_{m}(\alpha_{m}+1)\varGamma_{s}^{2}}{(8u_{B}+1)\alpha_{m}^{2}+2(2u_{B}+1)\alpha_{m}+1}\sqrt{u_{B}}\, . \label{55}
\end{equation}

An analytical expression for the shocked plasma's comoving electric field $E_{2}^{co}$ can be derived if we substitute $\beta_{2}=\sqrt{1-\dfrac{1}{\gamma_{2}^{2}}}$ in Eq. \ref{30}, use Eq. \ref{50} for $\gamma_{2}$ and then expand the resulting expression to $\order{\varGamma_{s}^{2}}$. The expression for $E_{2}^{co}$ obtained in this fashion is:

\begin{equation}
    \dfrac{E_{2}^{co}}{\sqrt{8\pi\epsilon_{1}}}=-\sqrt{\dfrac{u_{B}}{2(8u_{B}+1)\alpha_{m}^{2}+4(2u_{B}+1)\alpha_{m}+2}}\varGamma_{s}\, . \label{56}
\end{equation}

\section{Ideal MHD jump conditions}\label{appb}

The ideal MHD jump conditions are:
\begin{equation}
    \gamma_{2}n_{2}(\beta_{s}-\beta_{2})=n_{1}\beta_{s}\, , \label{a1}
\end{equation}
\begin{equation}
\begin{split}
    &\gamma_{2}^{2}(\epsilon_{2}+P_{2})(\beta_{s}-\beta_{2})-P_{2}\beta_{s}+\dfrac{(\beta_{2}^{2}+1)B_{2}^{2}}{8\pi}\beta_{s}-\dfrac{\beta_{2}B_{2}^{2}}{4\pi}=\\&      \epsilon_{1}\beta_{s}+\dfrac{B_{1}^{2}}{8\pi}\beta_{s}\, , \label{a2}
\end{split}
\end{equation}
\begin{equation}
    P_{2}-\gamma_{2}^{2}(\epsilon_{2}+P_{2})(\beta_{s}-\beta_{2})\beta_{2}+\dfrac{(\beta_{2}^{2}+1)B_{2}^{2}}{8\pi}-\dfrac{\beta_{2}B_{2}^{2}}{4\pi}\beta_{s}=\dfrac{B_{1}^{2}}{8\pi}\, , \label{a3}
\end{equation}
\begin{equation}
    (\beta_{s}-\beta_{2})B_{2}=\beta_{s}B_{1}\, . \label{a4}
\end{equation}

We follow the method described in Sect. \ref{sub2} and derive the following approximate analytical expressions for the shocked plasma:
\begin{equation}
    \gamma_{2}^{2}=\dfrac{\varGamma_{s}^{2}}{2+8u_{B}}\, , \label{a5}
\end{equation}
\begin{equation}
    r=\dfrac{2\varGamma_{s}^{2}}{1+8u_{B}}\, , \label{a6}
\end{equation}
\begin{equation}
    \dfrac{P_{2}}{\epsilon_{1}}=\dfrac{32u_{B}^{2}+12u_{B}+2}{(1+8u_{B})^{2}(3+8u_{B})}\varGamma_{s}^{2}\, , \label{a7}
\end{equation}
\begin{equation}
   \dfrac{ B_{2}}{\sqrt{8\pi\epsilon_{1}}}=\dfrac{2\sqrt{u_{B}}}{1+8u_{B}}\varGamma_{s}^{2}\, , \label{a8}
\end{equation}
which constitute simplified versions of the expressions presented in \citet{lyutikov2002}.
For $u_{B}\ll1$ the above expressions reduce to those for strongly relativistic shock propagation in non-magnetized media \citep{bm1976}:
\begin{equation}
    \gamma_{2}^{2}=\dfrac{\varGamma_{s}^{2}}{2}\, ,
\end{equation}
\begin{equation}
    r=2\varGamma_{s}^{2}\, ,
\end{equation}
\begin{equation}
    P_{2}=\dfrac{2\varGamma_{s}^{2}}{3}\epsilon_{1}\, ,
\end{equation}
while the shocked plasma's magnetic field satisfies:
\begin{equation}
    \dfrac{B_{2}}{\sqrt{8\pi\epsilon_{1}}}=2\sqrt{u_{B}}\varGamma_{s}^{2}\, .
\end{equation}
For $u_{B}\gg1$, these expressions give the following:
\begin{equation}
    \gamma_{2}^{2}=\dfrac{\varGamma_{s}^{2}}{8u_{B}}\, ,
\end{equation}
\begin{equation}
    r=\dfrac{\varGamma_{s}^{2}}{4u_{B}}\, ,\, ,
\end{equation}
\begin{equation}
    P_{2}=\dfrac{\varGamma_{s}^{2}}{16u_{B}}\epsilon_{1}\, ,
\end{equation}
\begin{equation}
    \dfrac{B_{2}}{\sqrt{8\pi\epsilon_{1}}}=\dfrac{\varGamma_{s}^{2}}{4\sqrt{u_{B}}}\, .
\end{equation}
\end{appendix}

\end{document}